\documentclass[12pt]{iopart}

\usepackage[aux]{rerunfilecheck}

\usepackage{commath}
\usepackage{mathtools}

\usepackage[style=english]{csquotes}

\usepackage[
  separate-uncertainty=true,   
  per-mode=symbol-or-fraction, 
]{siunitx}

\usepackage{xfrac}

\usepackage[
  labelfont=bf,        
  font=small,          
  width=0.9\textwidth, 
]{caption}
\usepackage{subcaption}

\usepackage{graphicx}
\usepackage{grffile}

\usepackage{float}
\usepackage{tabularx}

\usepackage{booktabs}

\usepackage[
  unicode,
  colorlinks=true,
  pdfcreator={},  
  pdfproducer={}, 
]{hyperref}

\usepackage{xcolor}
\hypersetup{
    colorlinks,
    linkcolor={green!30!black},
    citecolor={green!30!black},
    urlcolor={green!30!black}
}
\usepackage[shortcuts]{extdash}

\usepackage{icomma}
\usepackage{wrapfig}
\usepackage{rotating}
\DeclareSIUnit\annum{a}
\usepackage{pdfpages}
\usepackage{todonotes}
\usepackage{mdframed}
\usepackage{colortbl}
\usepackage{listings}

\usepackage[below]{placeins}

\usepackage[]{romannum}
\AtBeginDocument{\pagenumbering{arabic}}

\usepackage{chemformula}
\usepackage{duckuments}
\usepackage{todonotes}
\usepackage[section]{glossaries}
\usepackage[export]{adjustbox}

\usepackage{layouts} 

\usepackage[backend=biber,maxcitenames=2,
 mincitenames=1,
 maxbibnames=2,
 minbibnames=1,url=false,style=ext-authoryear-ibid,giveninits=true,date=year,doi=false,isbn=false,
 uniquelist=minyear, uniquelist=false, 
dashed=false
 ]{biblatex}
\DeclareDelimFormat{newunitpunct}{\space} 
\renewrobustcmd*{\bibinitperiod}{}
\DeclareNameAlias{sortname}{family-given}

\DefineBibliographyExtras{english}{}
\DeclareFieldFormat{biblabeldate}{#1}
\DeclareDelimFormat[bib]{nameyeardelim}{\space}
\DeclareDelimFormat[bib]{nametitledelim}{\addspace}
\DeclareDelimFormat[bib]{newunitpunct}{\addspace}
\renewbibmacro{in:}{} 

\DeclareFieldFormat[article,periodical,inproceedings]{volume}{\mkbibbold{#1}\nopunct}
\DeclareFieldFormat[article,periodical,inproceedings]{number}{}
\DeclareFieldFormat[article]{pages}{\mkcomprange{#1}\nopunct} 

\DefineBibliographyStrings{english}{
  page = {p}, 
  pages = {pp} 
}
\DeclareFieldFormat[book, inproceedings]{pages}{p\space\mkcomprange{#1} \nopunct}
\DeclareFieldFormat[inbook,incollection,inproceedings]{pages}{pp\space \mkcomprange{#1}\nopunct}  

\newbibmacro{string+doi}[1]{%
  \iffieldundef{doi}{\iffieldundef{url}{#1}{\href{\thefield{url}}{#1}}}{\href{http://dx.doi.org/\thefield{doi}}{#1}}}	

\DeclareFieldFormat[book]{title}{\usebibmacro{string+doi}{\mkbibemph{#1}}\nopunct}
\DeclareFieldFormat
  [article,inbook,inbook,inproceedings,patent,thesis,unpublished]
  {title}{\usebibmacro{string+doi}{#1\nopunct}}
\DeclareFieldFormat
  [suppbook,suppcollection,suppperiodical]
  {title}{\usebibmacro{string+doi}{#1}}

\DeclareListFormat[book,incollection]{location}{(#1}
\DeclareListFormat[book,incollection]{publisher}{#1)\nopunct}

\DeclareListFormat[inproceedings]{location}{#1\nopunct}
\DeclareListFormat[inproceedings]{publisher}{}

\DeclareFieldFormat{editortype}{\mkbibparens{#1}\nopunct}
\DefineBibliographyStrings{english}{%
  byeditor = {ed},
}
\DefineBibliographyStrings{english}{%
    andothers = {\em et\addabbrvspace al}
}

\DefineBibliographyExtras{english}{
  }
\DeclareListFormat[book,incollection, inproceedings]{editor}{#1\nopunct}

\begin{document}

\def\usepngs{1}

\title[]{Fast Maximum Likelihood Positioning for a Staggered Layer Scintillation PET Detector}

\author{C.~Lerche$^1$, W.~Bi$^{1,2}$, M.~Schöneck$^{1,\dagger}$, D.~Niekämper$^{1,2}$, Q.~Liu$^1$, E.~Pfaehler$^1$,  L.~Tellmann$^1$, J.~J.~Scheins$^1$, and N.~J.~Shah$^{1,3,4,5}$\\}

\address{$^1$ Institute for Neuroscience and Medicine (INM-4), Forschungszentrum Jülich GmbH, Germany\\
$^2$ Department of Physics, RWTH Aachen University, Aachen, Germany\\
$^3$ Institute of Neuroscience and Medicine 11 (INM-11), Forschungszentrum Jülich GmbH, Jülich, Germany \\
$^4$ JARA - BRAIN - Translational Medicine \\
$^5$ Department of Neurology, RWTH Aachen University, Aachen, Germany\\
$^\dagger$  M.~Schöneck is now with the Faculty of Medicine and Institute for Diagnostic and Interventional Radiology, University Hospital Cologne, University of Cologne, Cologne, Germany
\vspace{0.2cm}}
\ead{c.lerche@fz-juelich.de}


\begin{abstract}
In this study, we propose a fast implementation of a Maximum Likelihood Positioning (MLP) algorithm to estimate the energy and identify the active scintillator pixel in staggered layer scintillation detectors for PET. The staggered layer design with pixelated scintillators enables the determination of the gamma's depth of interaction and facilitates an iteration-free formulation of the MLP algorithm. The efficacy of the algorithm optimization was tested on a scintillation detector block designed for an ultra-high field BrainPET 7T, comprising three scintillator pixel layers. The three layers contain 24 × 24, 24 × 23 and 23 × 22 scintillator pixels, respectively, with a pixel pitch of 2 mm in both directions and layer thicknesses of 9, 8 and 7 mm. Calibration measurements, in combination with an automated calibration script, were used to obtain the expected counts of scintillation photons required in the MLP algorithm.  Using Single-Instruction-Multiple-Data parallelization, multi-threading and optimized cache lines, a maximum processing speed of approximately 22.5 million singles per second was achieved on a platform with four Intel Xeon Platinum 8168 CPUs and 60 threads, encompassing all required processing steps (i.e.~input/output, unpacking, timestamp extraction and calibration, single event clustering, MLP, storage). The automatic calibration failed for 1 to 15 individual scintillator pixels in approximately 10\% of the 120 scintillation detector blocks, necessitating manual correction. After applying the energy correction to the positioned single events, an energy resolution of $\Delta E=12\%\pm2\%\, \mathrm{FWHM}$ was obtained for the entire scintillation block. This value is very close to the energy resolutions measured for the individual scintillator pixels, proving that the MLP accurately identifies the scintillating pixel and that the energy correction method effectively compensates for the light collection variations of the SiPM array. 
\end{abstract}

\vspace{2pc}

\submitto{\PMB}
%
%
%

\section{Introduction}

Scintillation detectors are key components of gamma cameras and positron emission tomographs. A large variety of possible realizations using photomultiplier tubes (PMTs), avalanche photodiodes (APDs) or silicon photomultipliers (SiPMs) in combination with monolithic, semi-monolithic or pixelated scintillators of different materials have been proposed over the years. The output signals from the photodetectors are fed into an attached electronic and computing device to extract the photo conversion position, the energy of the converted gamma photon and the time of photo conversion. As with the design of scintillation detectors, there is a wide variety of methods for extracting these parameters, including centroid-based algorithms, also known as Anger logic, least-square methods and methods from statistical estimation theory, such as Maximum Likelihood Estimation (MLE) \parencite{Joung2001InvestigationCameras,Barrett2009Maximum-likelihoodDetectors,VanDam2011ImprovedDetectors,Liu2013ImprovedAlgorithm}. Perhaps the most important property of all these signal processing algorithms is processing speed since the number of scintillation events can easily go beyond 1 million counts per second per detector block in modern PET scanners. Accuracy, precision and robustness are also essential. Compared to other scintillation event positioning methods, statistically-based methods, such as MLE, provide a correct statistical approach which also allows the integration of relevant physical aspects. In this way, simplifying and idealizing assumptions often made to achieve practical solutions can be omitted. Apart from the inherent consideration of the Poisson random nature of scintillation light detection, MLE allows the integration of models for scintillation light propagation, missing signals, Compton scattering and many other effects exhibited by real scintillation detectors \parencite{Barrett2009Maximum-likelihoodDetectors}.
Moreover, the likelihood value can be used as a measure of the estimate quality \parencite{Lodomez2012DevelopmentPET-detectors,lodomez2014AparatusEvaluation}. 
Ignoring the real-world effects in the positioning algorithm generally leads to spatial bias and inaccuracies in the determined scintillation event position and energy, necessitating further correction efforts. To address these challenges, maximum likelihood positioning (MLP) has been proposed continuously by various groups since the 1970s, especially for scintillation detectors with monolithic scintillators \parencite{Gray1976MaximumCameras,Fessler1991RobustCameras,Gagnon1993MaximumInteraction,Joung2000ImplementationCameras,Joung2001CMiCE:Scheme,Wenze2007MaximumExperiments,Miyaoka2008DesignCapability,Lerche2009MaximumMeasurement,Hesterman2010Maximum-likelihoodAlgorithm}.
In addition, MLP implementations for pixelated scintillation detectors have also been proposed by \parencite{Lerche2011MaximumDetectors, Lerche2016MaximumDetectors,Gross-Weege2016MaximumScanners,berker2018ScintillationEvent,berker2018ScintillationEvent}. However, 
all mentioned MLP methods for determining the energy and position of the scintillation events are based on iterative algorithms, making them very time-consuming and thus potentially impractical for processing scintillation events in a typical PET scan with limited computational resources in a reasonable time. An iteration-free formulation of the MLP algorithm is also of great interest for implementations in Field Programmable Gate Arrays (FPGA) as it takes advantage of accelerated data processing with dedicated circuitry and massively parallelized signal processing pipelines. Several FPGA implementations of MLP algorithms have been proposed \parencite{DeWitt2010DesignPositioning,Johnson-Williams2011DesignAlgorithm,Wang2016AnDetector}. 
In this work, we present an iteration-free MLP algorithm for pixelated scintillation detectors with depth of interaction (DOI) detection capabilities based on a staggered scintillation layer design \parencite{Ito2010ASystem} and the algorithm previously presented in \parencite{Lerche2016MaximumDetectors}.  
As in the original implementation, the result of the position determination is the scintillator pixel ID together with the calibrated gamma photon energy. Thus, the additional association of the continuous $x$ and $y$ centroid (also called Anger) positions with these scintillator pixel IDs using Voronoi tessellations or similar approaches  \parencite{Du1999CentroidalAlgorithms} can be avoided during normal data acquisition. Although the implementation of the presented MLP algorithm in FPGAs is not the focus of this work, all special requirements of the specific FPGA hardware relevant for such an implementation have been taken into account \parencite{Lerche2019PatentMethodDetectors}.

\section{Methods}
\label{sect:mehtods}

\subsection{Iteration-free ML implementation}

The log-likelihood of a gamma photon interacting in a single elementary process, i.e., photoelectric effect or Compton effect, and depositing the energy, $E_{\rm d}$, in the scintillator pixel, $i$, given that the scintillation photons ${\bf q}=(q_1,q_2,\ldots,q_{N^{PD}})$ were detected on the photodetectors $(1,2,\ldots,{N^{PD}})$, is given by
\begin{equation}
  \label{eq:loglik-1}
  \mathcal{L}_i\left(i,E_{\rm d}|{\bf q}\right)=\ln\prod_{j}^{N_{PD}}\frac{e^{-\bar{q}_{i,j}}(\bar{q}_{i,j})^{q_j}}{q_j!},
\end{equation}
as the detection process is a Poisson process. In eq.~\ref{eq:loglik-1}, $N_{PD}$ denotes the total number of individual photodetectors or photodetector elements~$j$ of the photodetector array. The total amount $Q_E=\sum_{j}^{N_{PD}}q_j$  of scintillation photons detected in all photodetectors $j$ is proportional to the energy deposition $E_{\rm d}$, the gain or photon detection efficiency (PDE) of the photodetector, and the light yield of the scintillator. The latter two are assumed to be constant during any measurement. $\bar{q}_{i,j}$ is the expected average amount of detected scintillation photons, i.e. avalanches, in the photodetector element~$j$ if a scintillation event occurred in scintillator pixel $i$.

To obtain an iteration-free MLP algorithm, we replace $\bar{q}_{i,j}$ by

\begin{equation}
     \begin{aligned}
    & \bar{q}_{i,j} = \bar{q}_{i}^m \hat{q}_{i,j} \\
    & \text{with} \\
    & \bar{q}_{i}^m := \max\;\bar{\bf{q}}_i \;\;\text{ and} \\  
    & \hat{q}_{i,j}:=\frac{\bar{q}_{i,j}}{\max\;\bar{\bf{q}}_i}
     \end{aligned}
     \label{eq:substitute}
\end{equation}
i.e. $\bar{q}_i^m$ is the maximum value of all $\bar{q}_{i,j}$ values of the same scintillator pixel $i$ and $\hat{q}_{i,j}$ is the maximum-normalized expected avalanche distribution.
We obtain the log-likelihood function 
\begin{equation}
  \label{eq:loglik-2}
  \mathcal{L}_i\left(i,E_{\rm d}|{\bf q}\right)\approx\sum_j^{N_{PD}}q_j\ln\left(q_i^m\hat{q}_{i,j}\right)-q_i^m\hat{q}_{i,j},
\end{equation}
where all constant terms have been omitted because only the {\it argumentum maximi} is required. Furthermore, we approximated the expected maximum $\bar{q}_i^m$ by the detected value $q_i^m$, which introduces statistical uncertainty caused by Poisson variations. 
Due to these variations, the approximation $\bar{q}_i^m\approx q_i^m$ leads to an expected and acceptable relative average deviation of 3\% to 6\% from the exact value $\bar{q}_i^m$ for the scintillation detector used in this study. Typically, 300 to 800 scintillation photons are detected on the SiPM with the maximum signal. The scintillator pixel~$i^{ML}$ most likely to be hit is the one with the largest $\mathcal{L}_i$ value:
\begin{equation}
  \label{eq:iml}
  i^{ML} = \underset{i}{\arg \max}\,{\mathcal{L}_i\left(i,E_{\rm d}|{\bf q}\right)}.
\end{equation}
By knowing which scintillator pixel $i$ most likely emitted the light, the detected light can be corrected for variations in scintillation light detection efficiency due to boundary effects, insensitive photodetector areas, and individual trigger probabilities of the photodetector pixels. We can then convert the scintillation photons to energy:
\begin{equation}
  \begin{aligned}
 & E_{\gamma , i}^{ML} =  511\, \mathrm{keV}\frac{\sum_{j}^{N_{PD}}q_j}{\sum_{j:q_j>0}\bar{q}_{i,j}} = \frac{511\, \mathrm{keV}}{\bar{q}_{i^{ML}}^m}\frac{\sum_{j}^{N_{PD}}q_j}{\sum_{j:q_j>0}\hat{q}_{i^{ML},j}}.\\
& \text{with} \\ & \bar{q}_{i^{ML}}^m:= \max\;\bar{\bf{q}}_{i^{ML}}
\end{aligned}
\label{eq:e-correction}
\end{equation}
i.e. $\bar{q}_{i^{ML}}^m$ is the maximum of the most likely averaged signal vector, i.e. the maximum value of all values $\bar{q}_{{i^{ML}},j}$ with the same index $i^{ML}$.

\subsection{Scintillation detector and PET scanner}

\begin{figure}[t!]
  \centering
  $\vcenter{\hbox{
  \begin{subfigure}{0.45\textwidth}
  \includegraphics[interpolate=true,width=1\columnwidth]{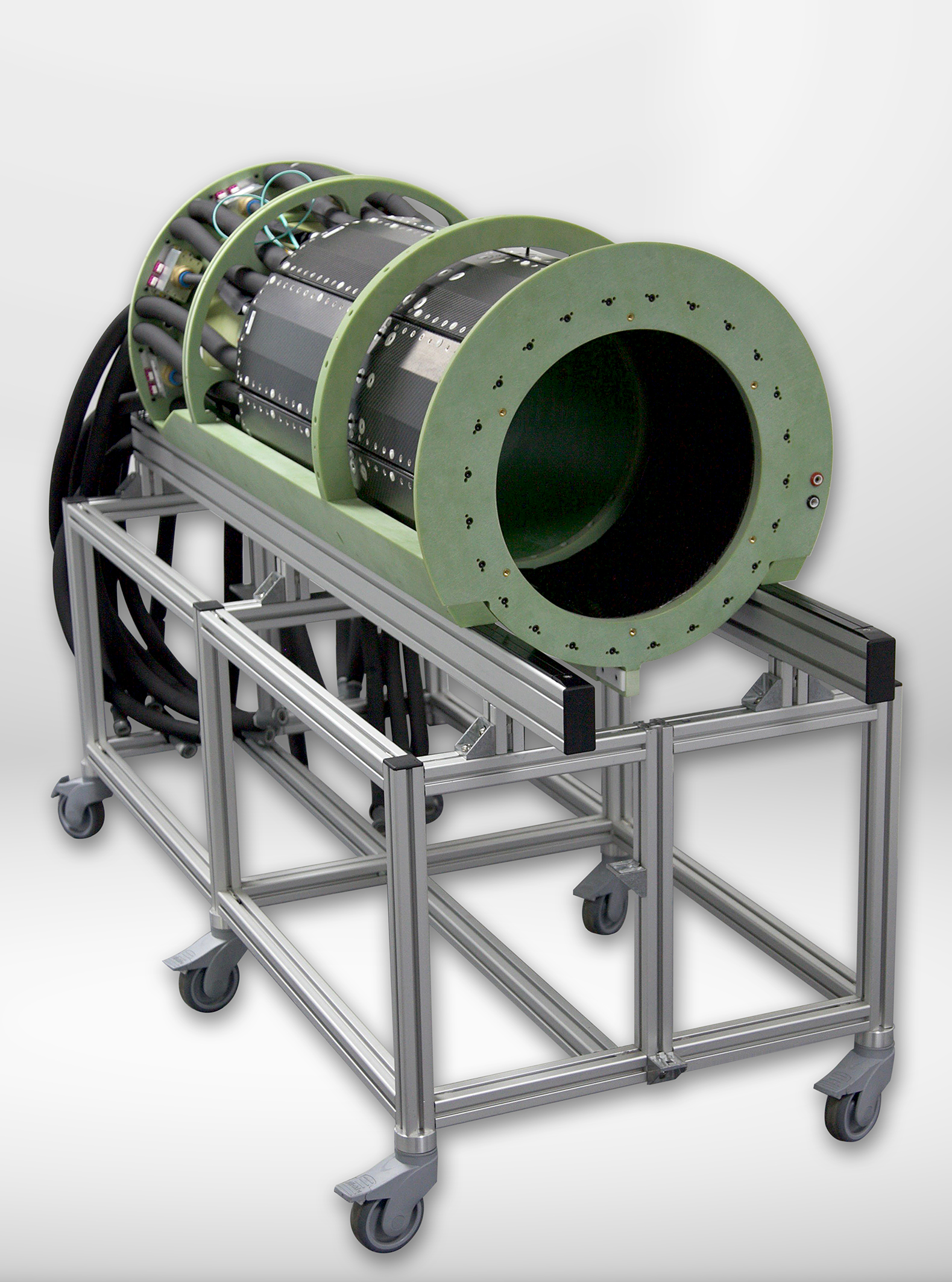}
  \caption{BrainPET 7T insert without cover and RF coil array on installation cart.}
  \label{subfig:PET-insert}
  \end{subfigure}}}$\hspace{3em}
  $\vcenter{\hbox{
  \begin{subfigure}{0.33\textwidth}
  \begin{subfigure}{1\columnwidth}
  \includegraphics[interpolate=true,width=1\columnwidth]{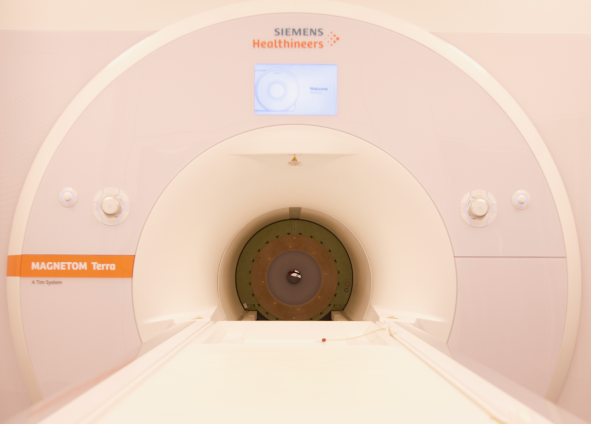}
  \caption{Front view of the BrainPET 7T with RF coil array inserted into a Siemens Terra Magnetom 7T MR system.}
  \label{subfig:PET-MR-System}
  \end{subfigure}\\[1eX]
  \begin{subfigure}{1\columnwidth}
  \centering
  \includegraphics[interpolate=true,width=0.86\columnwidth]{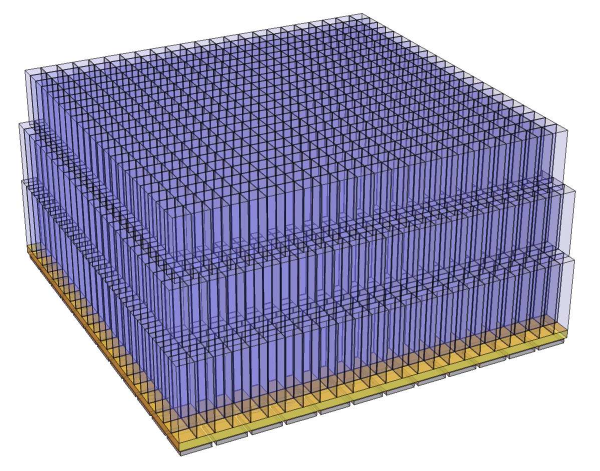}
  \caption{Schematic view of the staggered layer scintillation detector.}
  \label{subfig:scintillation-detetctor}
  \end{subfigure}\\[1eX]
  \begin{subfigure}{1\columnwidth}
  \centering
  \includegraphics[interpolate=true,width=0.86\columnwidth]{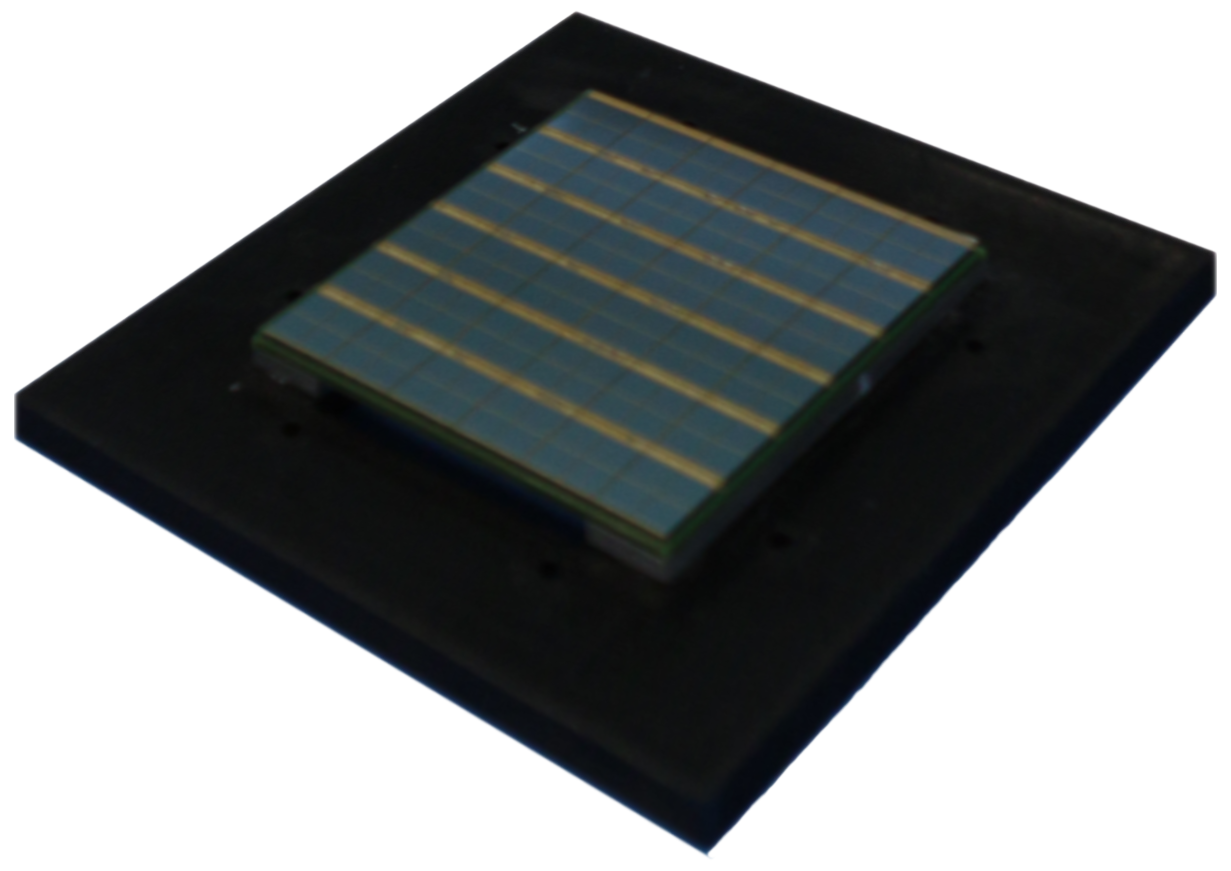}
  \caption{$\mathrm{12\times12}$ digital SiPM array.}
  \label{subfig:Sipm-array}
  \end{subfigure} 
  \end{subfigure}}}$
  \caption{BrainPET 7T insert, 7T MRI scanner with insert, sketch of the scintillation detector block and SiPM array.}
  \label{fig:PET-system}
\end{figure}

Although its use is not limited to this specific system, the improved ML positioning method has been developed for an ultra-high field (UHF) \mbox{BrainPET 7T} insert (fig.~\ref{subfig:PET-insert}) designed and built for neuro-scientific studies with a Siemens 7T human MRI scanner (MAGNETOM Terra, fig.~\ref{subfig:PET-MR-System}).
The BrainPET 7T insert consists of 8 PET modules, with each PET module housing $\mathrm{3\times 5}$ scintillation detector blocks (in total 120) and the associated digital electronics \parencite{Weissler2020HyperionPET/MRI,Lerche2023FirstNeuroscience,Lerche2024MRApplications}. A pixelated scintillation detector design with three staggered layers of LSO was chosen to provide depth of interaction detection and homogeneous spatial image resolution across the entire PET FOV \parencite{Ito2010ASystem, Lenz2021DesignStudies}. The three scintillator layers contain $\mathrm{24\times 24}$,  $\mathrm{24\times 23}$ and $\mathrm{23\times 22}$  scintillator pixels, respectively, with a pixel pitch of 2~mm in both directions and layer thicknesses of 9, 8 and 7 mm. Each of the 1634 individual pixels is wrapped in 3M ESR reflector, and the scintillator array is coupled via a 1 mm thick light guide to an array of $\mathrm{12\times12}$ digital SiPMs DPC-3200 from Philips Digital Photon Counting \parencite{Frach2009ThePerformance}. 
All scintillator surfaces, except the one coupled to the SiPM matrix, were roughened by grinding. The edges of the light guide were covered with light-absorbing material. The pitch of the SiPMs is exactly 4~mm in both directions, and the (UHF MR compatible) array of SiPMs was designed and manufactured by Hyperion Imaging Systems GmbH, Aachen, Germany \parencite{Weissler2020HyperionPET/MRI}. The first (lowest) layer of scintillator pixels is aligned with the SiPM array such that exactly $\mathrm{2\times2}$ scintillator pixels are positioned exactly over one of the $\mathrm{12\times12}$ SiPMs. The middle layer of scintillator pixels is offset by half of the scintillator pixel pitch in the transverse direction of the PET FOV with respect to the lowest layer. The uppermost layer of scintillator pixels is offset with respect to the middle layer by half of the scintillator pixel pitch in the transverse and axial direction with respect to the PET's FOV. These offsets give rise to a characteristic pattern when centroid positions are calculated, which enables the reliable identification of the layer \parencite{Ito2010ASystem,Lenz2021DesignStudies}.

\subsection{Calibration}
\label{sec_calib}
 To determine the log-likelihood according to eq.~\ref{eq:loglik-2} for each of the scintillation events during a normal measurement, the maximum-normalized expected avalanche count distribution $\hat{q}_{i,j}$ and the expected average amount of detected scintillation photons, (avalanches) $\bar{q}_{i,j}$ must be determined in advance. The latter is obtained from a calibration measurement. For these measurements, three $^{68}\mathrm{Ge}$ line sources with an approximate total activity of 30~MBq were placed in the FOV of the PET scanner. The scan time was adjusted to collect approximately 5 million scintillation events (singles) for each of the 120 scintillation detector blocks. All operating parameters of the digital SiPM (overvoltage, trigger setting, validation time, integration time), except the trigger validation network, were set to the same values as for normal operations. The trigger validation network was set to \texttt{0x55:OR}, which corresponds to an average threshold of $\mathrm{16.9\pm6.2}$ detected scintillation photons to accept an SiPM hit \parencite{Schug2016InitialTechnology}. For normal operations, the trigger validation network is set to \texttt{0x50:OR}, corresponding to an average validation threshold of $\mathrm{37.1\pm12.8}$ scintillation photons. The different settings for the trigger validation network during calibration and normal operation were intentionally chosen to assure a sufficiently accurate sampling of the signal distribution during calibration and reduced data throughput during normal operation. The pre-processing of the calibration measurement data includes, in this order: 
 \begin{itemize}
     \item Temporal clustering of the SiPM hits with a clustering window of $\mathrm{20~ns}$, 
     \item Suppression of all events with an incomplete neighbourhood of $\mathrm{3\times3}$ triggered SiPMs,
     \item Computation of the centroid position of the complete $\mathrm{3\times3}$ neighbourhood, and
     \item Suppression of all events where the sum of all triggered SiPM values $\sum_jq_j$ is less than 50, greater than 4000 and exceeds the sum of the SiPM values of the $\mathrm{3\times3}$ neighbourhood by a factor of 1.7 to filter out pulse pile-up (PU) events. 
 \end{itemize}
 Peaks were then extracted from the 2D histogram, e.g.~fig.~\ref{subfig:anger-2D-flood} for all $x$ and $y$ centroid positions corresponding to the regular grid of scintillator pixel positions (the data is first split into sub-histograms for individual SiPM rows and columns to increase the robustness of the algorithm). Due to a low gain variation of less than 20\% between SiPMs on the same sensor, the individual scintillator pixels provide a very good estimate of the expected centroid positions calculated from the scintillation events. Using the centroid positions, the dataset was then divided into 144 independent datasets corresponding to the 144 SiPMs of the array. The peak positions extracted from the $x$ and $y$ position histograms were used to generate a grid of intervals whose centroid positions correspond to the individual scintillator pixels.   
 Next, the sum of the SiPM values of the $\mathrm{3\times3}$ neighbourhood of all events falling within a single 2D interval corresponding to a specific scintillator pixel was binned into a histogram and fitted with a function comprising a Gaussian function and an empirical background term ($E_{A, 9}^\kappa e^{-\lambda E_{A,9}}$, where $E_{A,9}$ is the sum of the SiPM values of the $\mathrm{3\times3}$ neighbourhood and $\kappa$ and $\lambda$ are free fit parameters). For a few cases, this generic generation of 2D grid intervals is not accurate enough due to larger distortions in the 2D centroid position histogram than in the majority of cases. However, these cases can be identified by the number of low event counts in the 2D interval and are then manually corrected. 
 
 Due to the known boundary effects of the centroid positioning algorithm, the centroid positions of  adjacent scintillator pixels in different layers (exclusively at horizontal/vertical borders and corners) overlap and cannot be separated by the centroid position alone (see fig.~\ref{subfig:cluster-by-layer}). However, due to variations in the light collection efficiencies of different scintillator pixel layers, the energy spectrum of events falling inside these 2D grid intervals exhibit two completely separated full absorption peaks (see fig.~\ref{subfig:pixel-e-hists-double-peak}). This allows the separation of the data sets from the different pixels \parencite{Sampath2013CharacterizationApplications,Lerche2014PixelArrays,Lenz2021DesignStudies}. For these special cases, the fit model consists of two Gaussian functions and the same background model as described previously. All signal values for events falling within the total avalanche count interval $[-3\sigma, 3\sigma]$ are selected to compute the 144 average SiPM values $\hat{q}_{i,j}$ for all 1634 scintillator pixels. For the least-squares fitting of the energy histograms of the individual scintillator pixels, it was extremely important to ensure good initial parameters in order to achieve reliable convergence in the fitting process. For this purpose, the starting parameters for the photopeak positions were determined using a fixed quantile, and the starting parameters for the background and peak amplitudes were determined using fixed fractions of the total spectrum counts. The remaining fit start parameters were set to the same value for all pixels. All calibration steps were implemented as an automatically running script using Mathematica version 12.1.1.0.  

\subsection{Algorithm implementation}

The iteration-free formulation of the MLP algorithm is one of several measures taken to increase the processing speed and data throughput of the algorithm without sacrificing positioning accuracy. For a typical scintillation event, it is intuitively clear that very few of the 1634 scintillator pixels are possible candidates for obtaining a maximum according to equations \ref{eq:iml} and \ref{eq:loglik-2}. When a signal $\bf{q}$ is detected, the potentially scintillating pixel must be closely located above one of the triggered SiPMs, unless this SiPM is not yet recharged after a previous event, i.e. during dead time. The latter is very unlikely due to the compact dimensions of the SiPMs. For most scintillator pixels, the log-likelihood will be very small because they are located over inactive SiPMs. The minimum candidate group of scintillator pixels that should be included in the ML estimation was determined in a Monte Carlo (MC) simulation with GATE and with optical photon transport for a scintillation detector. In the GATE simulation, all relevant design parameters, i.e. pixel pitch, light guide design, number and thickness of layers, and digital SiPM behaviour, were the same as in the original scanner. Only the number of SiPMs was reduced from $\mathrm{12\times12}$ to $\mathrm{4\times4}$, since the detector behaviour is translation invariant over a large part of the photodetector area \parencite{Lenz2021DesignStudies}. When including more than 16 scintillator pixels ($\mathrm{2\times2}$ over one SiPM in the bottom layer, $\mathrm{2\times3}$ and $\mathrm{3\times2}$ centred over the same SiPM in the middle and top layers, respectively), the ML positioning accuracy reached approx. 70\% and did not increase further\footnote{It has been shown in \parencite{Lenz2021DesignStudies} that, due to inter-crystal scattering, a positioning accuracy greater than 77\% cannot be achieved for the detector design used.}. A candidate look-up table was generated containing the 16 relevant scintillator pixels for each of the 144 SiPMs. In the same way, the minimum required number of included SiPM pixels was determined to be 9. 

Another important approach for optimizing the speed of the MLP algorithm is to match the accuracy of the computation to the accuracy of the underlying measurement data. In the case of scintillation detectors, the measurement precision is limited by the Poisson-nature of photon counting. For a typical detected scintillation event $\bf{q}$, the Poisson uncertainty for the value $q_m$ is about 3\%, and about 25 \% for the value just above the mean trigger threshold. In contrast, the machine accuracy for single precision float variables is about 1 ppm. We can, therefore, reduce the computational precision and use a faster, approximate implementation of the logarithm proposed by Paul Mineiro \parencite{mineiro2011FasApproximate}:
\begin{lstlisting}
inline float customLog(float arg) { 
    if(arg > 0.f) { 
        union { float f; uint32_t i; } vx = { float(arg) };
        float y = vx.i;
        y *= 1.0 / (1 << 23);
        return float(0.69314718f * (y - 126.94269504f));
    } else return 0.f; 
}
\end{lstlisting}
The case $\ln(0)=-\infty$, i.e. a vanishing argument of the logarithm, is excluded by setting $\ln(0)=0$, and thus suppressing a contribution of inactive SiPMs to the log-likelihood in eq.~\ref{eq:loglik-2}. 

The iteration-free formulation of the MLP allows an efficient implementation by using Single-Instruction-Multiple-Data (SIMD) Intel AVX intrinsics~\parencite{intel2021} in combination with optimized cache line access~\parencite{Kowarschik2003}. To achieve this, each scintillator pixel candidate group of 16 scintillator pixels $i$ addresses a tailored list of maximum-normalized, expected avalanche distribution vectors containing values $\hat{q}_{i,j}$ (re-)sorted by photodetector elements $j$. In this way, multiply-add operations can be performed as AVX instructions to calculate all log-likelihood values ${\cal{L}}_i$ simultaneously. The MLP processing is embedded into the full singles processing software, which involves the following subsequent steps:
\begin{itemize}
    \item Raw data decoding/unpacking of all acquired hits,
    \item Timestamp calculation (including skew corrections) for all hits, 
    \item Temporal sorting of hits and clustering (separately for each detector block), 
    \item MLP for all formed clusters, including energy calculation and
    \item Extracting/storing all singles by scintillator pixel index and calibrated energy.
\end{itemize}
The multi-thread data processing runs on a single machine with four Non-Unified Memory Access (NUMA) nodes driven by Ubuntu Linux (kernel version 5.4.0-204-generic). Each node is equipped with an Intel Xeon Platinum 8168 CPU @ 2.70 GHz (24 cores) and 96 GB/node of memory. Hence, in total, 96 cores are available for multi-threading while an optimized IO-performance was found for 60 threads in parallel. All C++ codes have been compiled using gcc 9.4.0 with optimization flags \texttt{-O3 -ffast-math}.

\section{Results}

\subsection{Calibration}

\begin{figure}[t!]
  \centering
  \begin{subfigure}{0.5\textwidth} \includegraphics[interpolate=true,width=1\columnwidth]{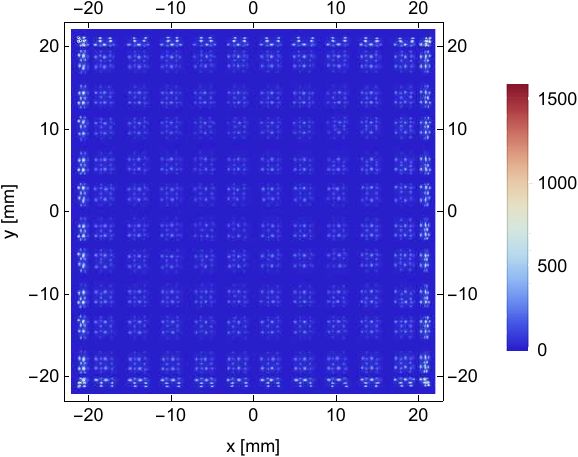}
  \caption{2D histogram of centroid positions for the scintillation events}
  \label{subfig:anger-2D-flood}
  \end{subfigure}\hspace{0em}   
  \begin{subfigure}{0.39\textwidth}  \includegraphics[interpolate=true,width=1\columnwidth]{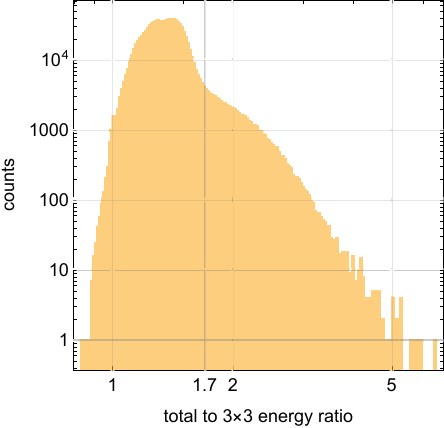}
  \caption{Ratio of the sum of all triggered SiPM values to the sum of $\mathrm{3\times3}$ neighbourhood.}
  \label{subfig:angereng-tot-eng-ratio}
  \end{subfigure}\\[1eX]
  \begin{subfigure}{0.9\textwidth} \includegraphics[interpolate=true,width=1\columnwidth]{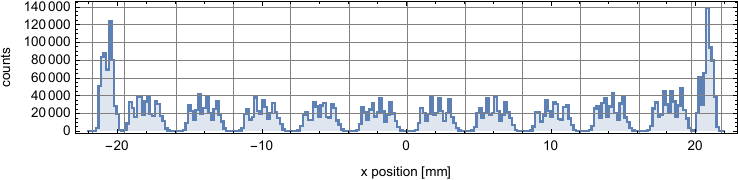}
  \caption{$x$ projection for the 2D histogram of centroid positions}
  \label{subfig:anger-flood-x-proj}
  \end{subfigure}\\[1eX]
  \begin{subfigure}{0.9\textwidth} \includegraphics[interpolate=true,width=1\columnwidth]{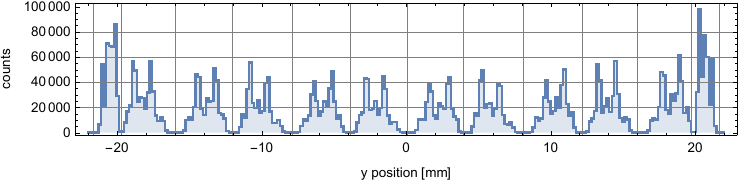}
  \caption{$y$ projection for the 2D histogram of centroid positions}
  \label{subfig:anger-flood-y-proj}
  \end{subfigure}
  \label{subfig:anger-positin0n-rersults}
  \caption{2D histogram of centroid positions and corresponding 1D projections of a single detector block as obtained by using the calibration measurement data.}
\end{figure}

Fig.~\ref{subfig:anger-2D-flood} shows the 2D histogram of the computed centroid positions for all registered calibration events for one of the 120 scintillation detector blocks (see section~\ref{sec_calib}) after suppression of pulse pile-up and events with avalanche counts of less than 50 and greater than 4000. Almost all of the 1634 centroid positions corresponding to individual scintillator pixels are resolved, except for those at the boundaries, which result in equal centroid positions but different amounts of light collection. The positions of the 144 individual SiPMs can also be recognized as the centroid positions of scintillator pixels above the same SiPM are closer together. Fig.~\ref{subfig:angereng-tot-eng-ratio} shows the ratio of the sum of all triggered SiPM values $\sum_jq_j$ to the sum of the SiPM values of the $\mathrm{3\times3}$ neighbourhood of the maximum SiPM, which is used to filter out pulse pile-up events. A clear change in the behaviour of the histogram can be observed at the value of 1.7, which is the threshold for pile-up events. Figs.~\ref{subfig:anger-flood-x-proj} and \ref{subfig:anger-flood-y-proj} show the 1D histograms of the same calculated centroid positions projected in $x$ and $y$ directions. The 11 deepest valleys in both histograms could be effectively used to segment the complete position map (fig.~\ref{subfig:anger-2D-flood}) into 144 individual 2D histograms corresponding to the events $\bf{q}$, whose maximum $q_m$ was detected with the respective SiPM. The amplitudes of the maxima in these 1D partial histograms for the individual SiPM rows and columns differed sufficiently from the amplitudes of the intervening minima to allow the reliable identification of the intervals for the centroid positions of the individual scintillator pixels in most cases.

\begin{figure}[t!]
  \centering
  \begin{subfigure}{0.4\textwidth}
  \if\usepngs1
  \includegraphics[interpolate=true,width=1\columnwidth]{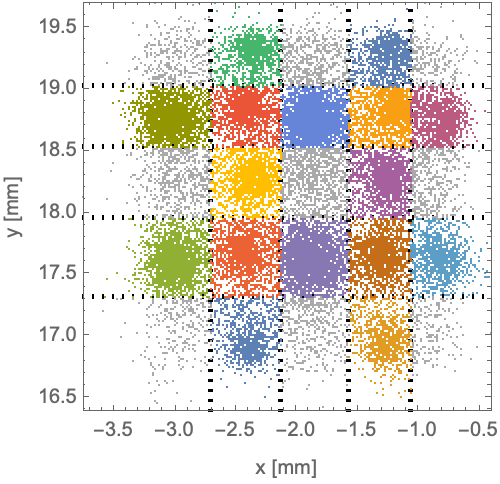}
    \else
  \includegraphics[interpolate=true,width=1\columnwidth]{figures/Fig-single-cent-SiPM-flood-auto.eps}
  \fi
  \caption{centre SiPM, automatic segmentation}
  \label{subfig:single-cent-sipm-flood-auto}
  \end{subfigure}\hspace{2em}
    \begin{subfigure}{0.4\textwidth}
  \if\usepngs1
    \includegraphics[interpolate=true,width=1\columnwidth]{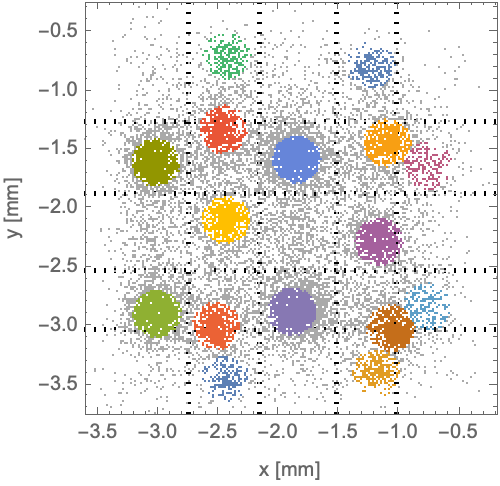}
    \else
    \includegraphics[interpolate=true,width=1\columnwidth]{figures/Fig-single-cent-SiPM-flood-man.eps}
  \fi
  \caption{centre SiPM, manual segmentation\\[3eX]}
  \label{subfig:single-cent-sipm-flood-man}
  \end{subfigure}\\[1eX]

  \begin{subfigure}{0.3\textwidth}
  \if\usepngs1
   \includegraphics[interpolate=true,width=1\columnwidth]{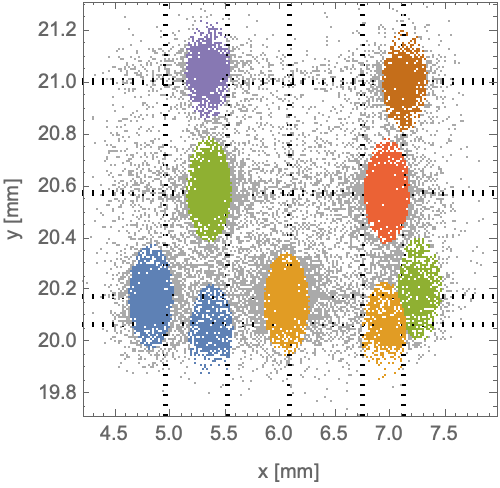}
   \else
   \includegraphics[interpolate=true,width=1\columnwidth]{figures/Fig-single-hor-SiPM-flood-man.eps}
 \fi
  \caption{horizontal border SiPM, manual segmentation}
  \label{subfig:single-hor-sipm-flood-man}
  \end{subfigure}\hspace{1em}
  \begin{subfigure}{0.3\textwidth}
  \if\usepngs1
  \includegraphics[interpolate=true,width=1\columnwidth]{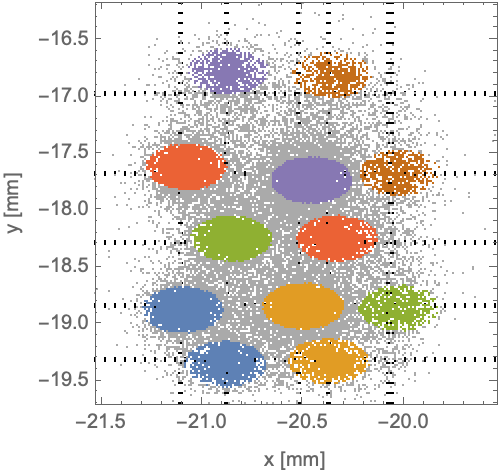}
    \else
  \includegraphics[interpolate=true,width=1\columnwidth]{figures/Fig-single-vert-SiPM-flood-man.eps}
  \fi
  \caption{vertical border SiPM, manual segmentation}
  \label{subfig:single-vert-sipm-flood-man}
  \end{subfigure}\hspace{1em}
 \begin{subfigure}{0.3\textwidth}
  \if\usepngs1
  \includegraphics[interpolate=true,width=1\columnwidth]{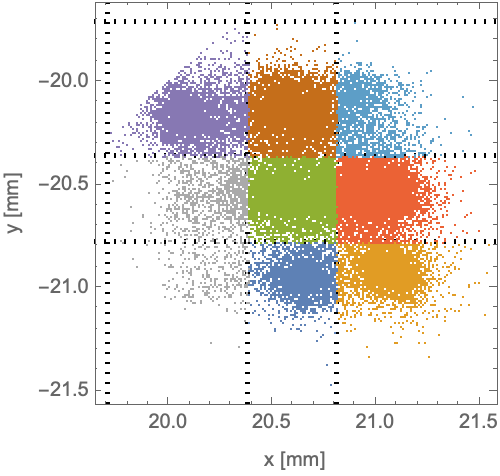}
    \else
  \includegraphics[interpolate=true,width=1\columnwidth]{figures/Fig-single-corner-SiPM-flood-auto.eps}
  \fi
  \caption{corner SiPM, automatic segmentation}
  \label{subfig:single-corner-sipm-flood-auto}
  \end{subfigure}\\[1ex]
   \begin{subfigure}{1\textwidth}
    \includegraphics[interpolate=true,width=1\columnwidth]{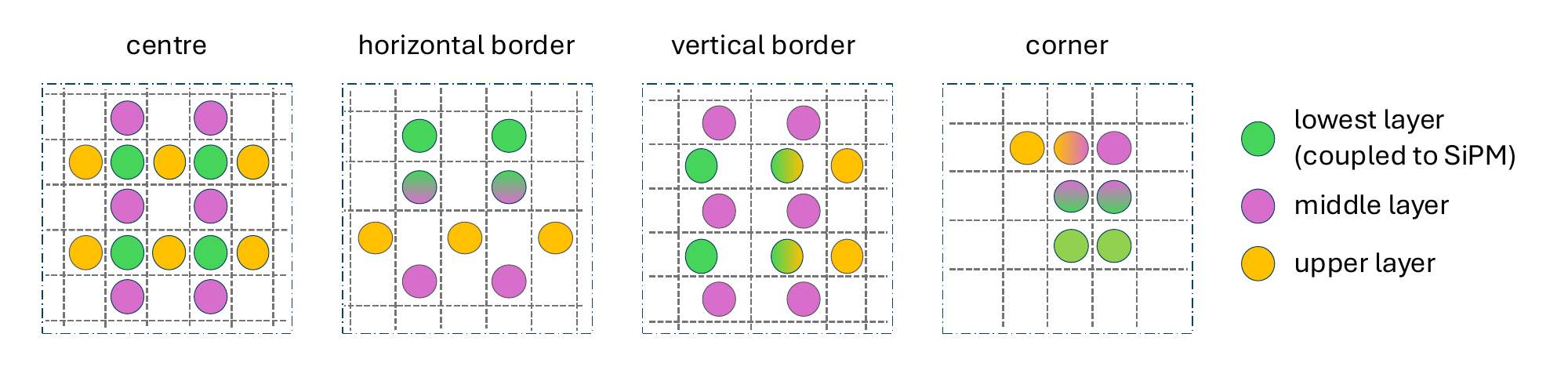}
    \caption{Illustration of which cluster belongs to which layer: Two-coloured circles represent clusters with overlapping centroid positions, i.e. clusters whose centroid positions belong to two layers.}
  \label{subfig:cluster-by-layer}
  \end{subfigure}
   \caption{Examples of clustered individual SiPM flood maps. The dotted grid lines in all 2D histograms represent the grid of 2D centroid position intervals for individual scintillator pixels derived from the maxima and minima of the 1D histograms shown in Figs.~\ref{subfig:anger-flood-x-proj} and \ref{subfig:anger-flood-y-proj}. Colours are used to highlight those events that fall into the 2D regions for further processing.} 
  \label{fig:single-sipm-floods}
\end{figure}

Examples of 2D histograms for individual SiPMs are shown in Fig.~\ref{fig:single-sipm-floods}. For about 10\% of the 120 scintillation detector blocks, the automatic calibration failed for 1 to $\approx$~15 individual scintillator pixels because the centroid positions of too few scintillation events fall within the 2D position intervals (determined globally from the 1D position histograms) for the corresponding scintillator pixels.  
\begin{figure}[t!]
  \centering
  \begin{subfigure}{0.4\textwidth}
  \begin{subfigure}{1\columnwidth} \includegraphics[interpolate=true,width=1\columnwidth]{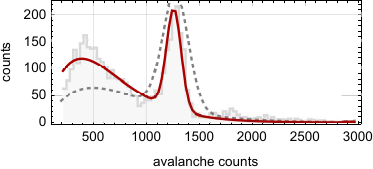}
  \caption{Case with a single photopeak spectrum.}
  \label{subfig:pixel-e-hists-single-peak}
  \end{subfigure}\\
  \begin{subfigure}{1\columnwidth} \includegraphics[interpolate=true,width=1\columnwidth]{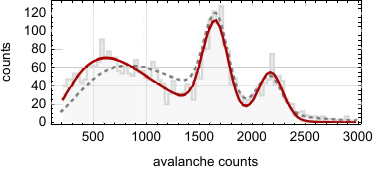}
  \caption{Case with a double photopeak spectrum.}
  \label{subfig:pixel-e-hists-double-peak}
  \end{subfigure}
  \end{subfigure}\hspace{0em}
  \begin{subfigure}{0.55\textwidth} \includegraphics[interpolate=true,width=1\columnwidth]{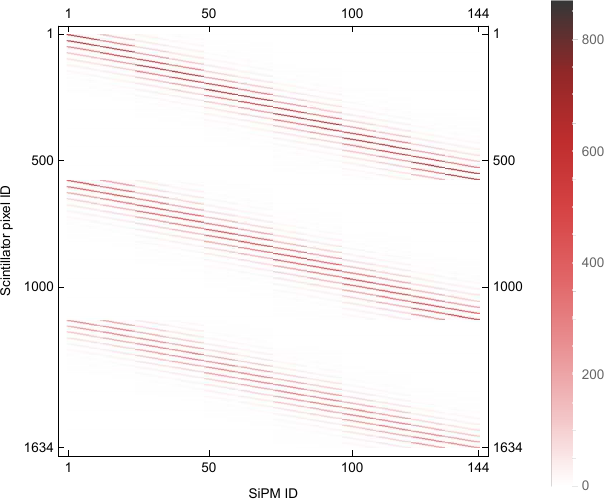}
  \caption{Measured average avalanche counts for all scintillation pixels and all SiPMs, (expectation value matrix)}
  \label{subfig:light-dist-matrix}
  \end{subfigure}\\[1eX] 
  \begin{subfigure}{1\textwidth} \includegraphics[interpolate=true,width=1\columnwidth]{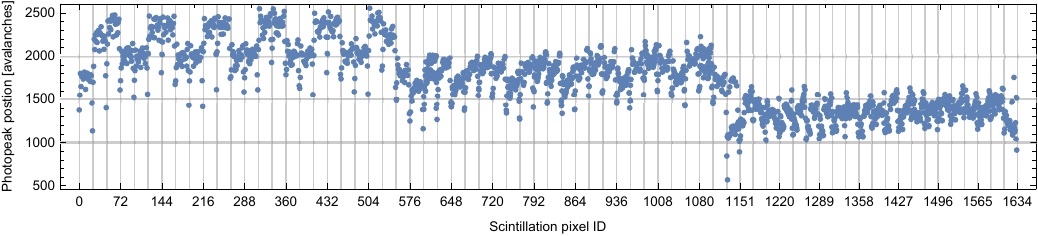}
  \caption{Photopeak positions for all 1634 individual scintillator pixels.}
  \label{subfig:photopeak-positions-all-xtals}
  \end{subfigure}\\[1eX]
  \begin{subfigure}{1\textwidth} \includegraphics[interpolate=true,width=1\columnwidth]{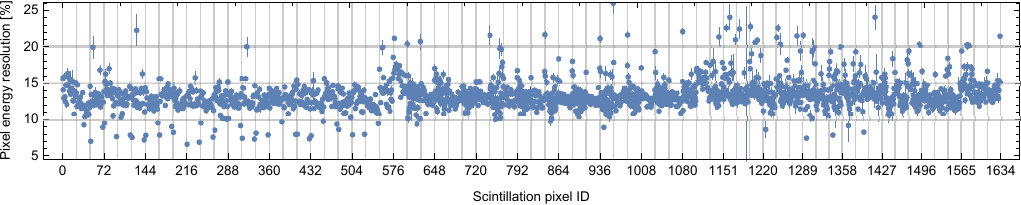}
  \caption{Pixel-wise energy resolutions for all 1634 individual scintillator pixels.}
  \label{subfig:photopeak-eres-all-xtals}
  \end{subfigure}\\[1eX]
  \begin{subfigure}{1\textwidth} \includegraphics[interpolate=true,width=1\columnwidth]{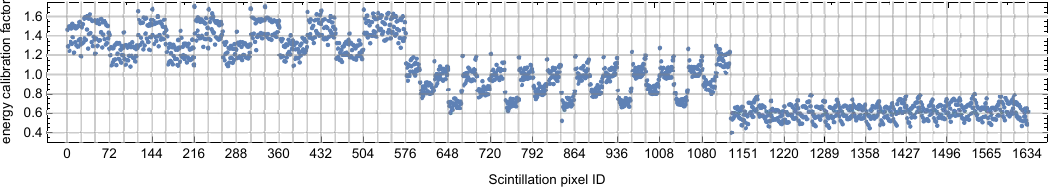}
  \caption{Calibration factors to convert from avalanches on SiPM with maximum signal to energy in keV.}
  \label{subfig:ecal-factors}
  \end{subfigure}
  \caption{Example energy histograms of a single scintillator pixel. Red lines show the best fit, dashed lines show the fit model with start parameters (\ref{subfig:pixel-e-hists-single-peak},\ref{subfig:pixel-e-hists-double-peak}); scintillation light distribution matrix (\ref{subfig:light-dist-matrix}); full absorption peak positions (\ref{subfig:photopeak-positions-all-xtals}); energy resolutions (\ref{subfig:photopeak-eres-all-xtals}); and matrix conversion factors (\ref{subfig:ecal-factors}).}  
  \label{fig:pixel-e-hists}
\end{figure}

\begin{figure}[t!]
  \centering
    \begin{subfigure}{1\textwidth}\centering
  \includegraphics[interpolate=true,width=0.85\textwidth]{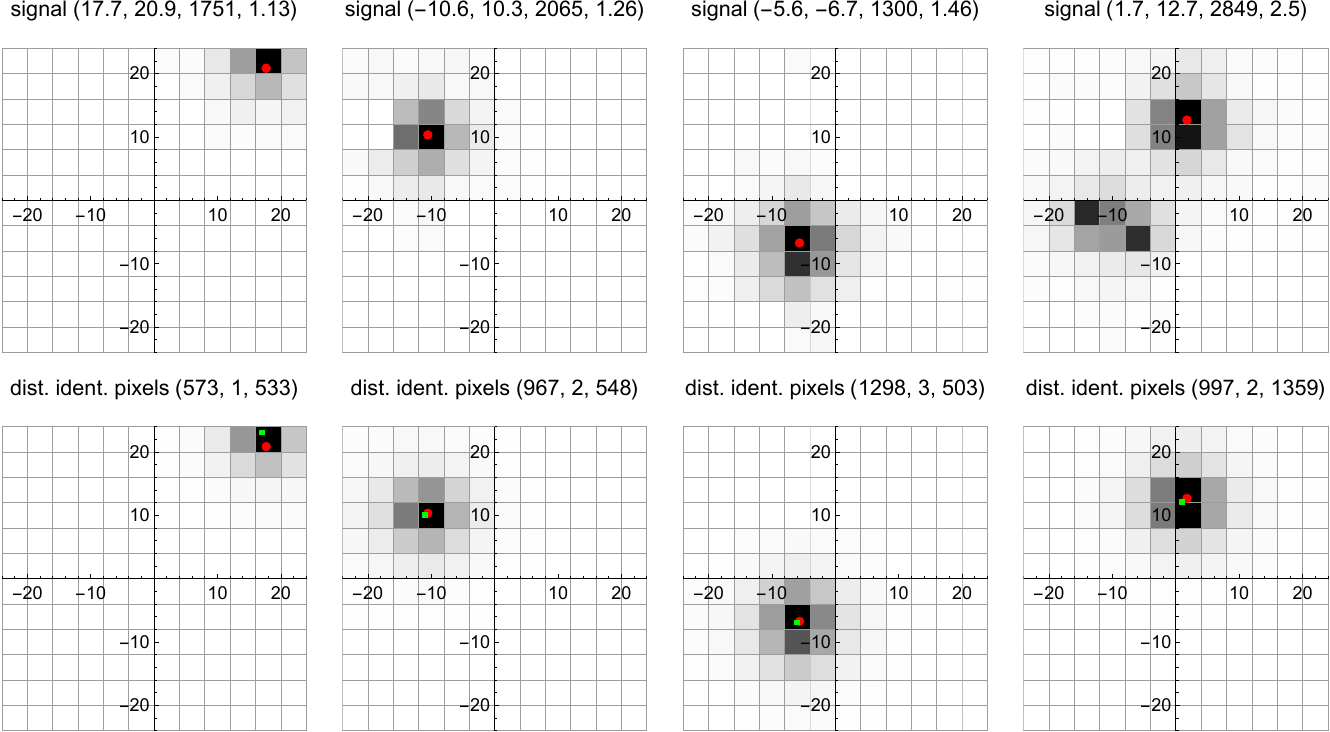}
  \caption{Detected signal and centroid position (top row) and ML position (bottom row) for four typical scintillation events (lowest, middle, upper layer, and pile-up event). The titles of the plots in the upper row contain the values of the centroid positions (in mm), the sum signal of the $\mathrm{3\times3}$ neighbourhood (uncorrected centroid energy), and the ratio of the sum of all triggered SiPM values $\sum_jq_j$ to the sum of the SiPM values of the $\mathrm{3\times3}$ neighbourhood. The titles of the plots in the bottom row contain the pixel ID, the layer ID and the corrected energy in keV. In the rightmost column, it can be observed that only one scintillation event is considered with the MLP algorithm.}
  \label{subfig:exemple-event-positions}
  \end{subfigure}\\[1eX]
  \begin{subfigure}{0.95\textwidth}
  \includegraphics[interpolate=true,width=1\textwidth]{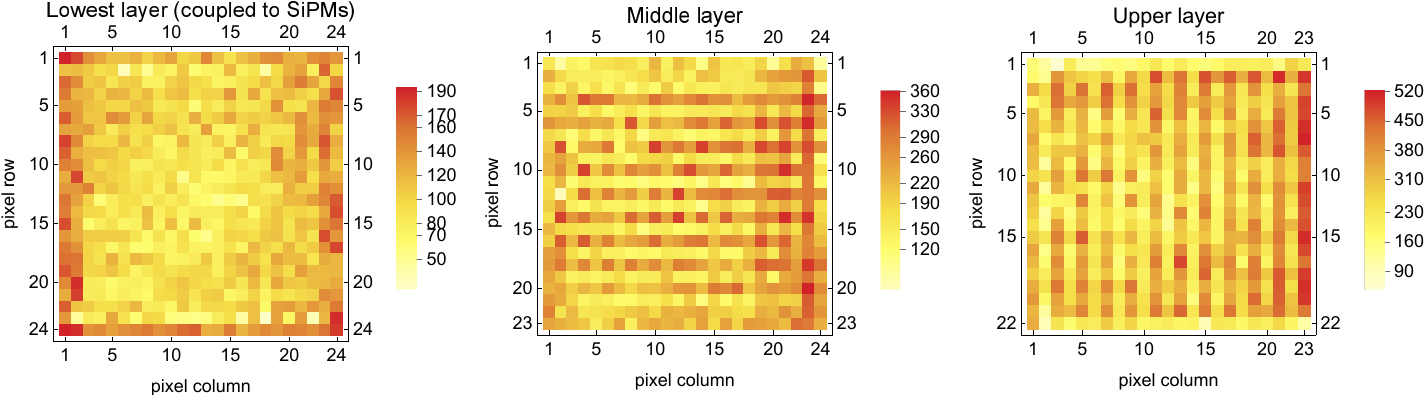}
  \caption{Scintillation event counts after ML positioning for each scintillator pixel.} 
  \label{subfig:xtal-counts}
  \end{subfigure}\\[1eX]
   \begin{subfigure}{0.43\textwidth}
    \includegraphics[interpolate=true,width=1\textwidth]{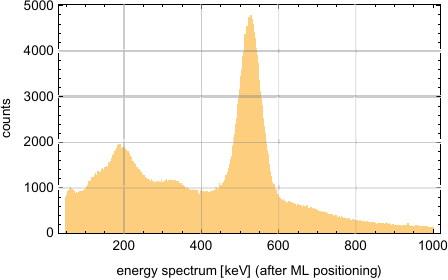}
  \caption{Energy histogram after ML positioning and energy correction.} 
  \label{subfig:energy-hist-MLP}
 \end{subfigure}\hspace{2em}
  \begin{subfigure}{0.43\textwidth}
    \includegraphics[interpolate=true,width=1\textwidth]{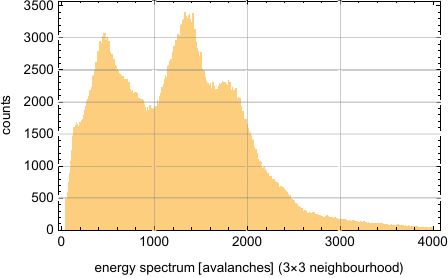}
  \caption{Energy histogram obtained with centroid positioning using $\mathrm{3\times3}$ neighbourhood.}
  \label{subfig:energy-hist-anger}
 \end{subfigure}
 \caption{Example events, position counts and energy spectra before and after applying the ML positioning algorithm.}
 \label{fig:MLP-results}
\end{figure}

For those SiPMs where the automatic calibration was not successful, the centroid points of different scintillator  pixels were manually determined, and the calibration results were combined with the automatically determined results. Histograms and best-fit results for the distribution of avalanche counts (sum of the neighbourhood of $\mathrm{3\times3}$ SiPMs) are shown in Fig.~\ref{subfig:pixel-e-hists-single-peak} and \ref{subfig:pixel-e-hists-double-peak} for two typical cases with a single full absorption peak and two full absorption peaks. 
Fig.~\ref{subfig:light-dist-matrix} shows the measured average avalanche counts for all 144 SiPMs and for all 1634 scintillator pixels. 
The first $\mathrm{24\times24}$ rows correspond to the lowest scintillator pixel layer, i.e.~the one bonded to the light guide. The last $\mathrm{23\times22}$ rows correspond to the uppermost scintillator pixel layer. The determined positions of the full absorption peak and the corresponding energy resolutions of the individual scintillation pixels are shown for all 1634 scintillator pixels in Fig.~\ref{subfig:photopeak-positions-all-xtals} and \ref{subfig:photopeak-eres-all-xtals}, respectively. A relevant modulation of the light collection efficiency due to insensitive areas of the SiPM array and edge effects can be seen. The different light collection efficiencies for the three layers are also clearly visible. The corresponding energy calibration factors $511 \mathrm{keV}/\bar{q}_{i^{ML}}^m$ are plotted in Fig.~\ref{subfig:ecal-factors}.

\subsection{ML positioning and energy correction}

Fig.~\ref{subfig:exemple-event-positions} shows some examples of positioning results for measured scintillation events corresponding to different scintillator pixel layers. The matrix plots in the top row show the detected avalanche counts of all $12\times 12$ SiPMs for four measured individual scintillation events. The corresponding centroid position is shown as a red dot in the same plot together with the signal distribution. The matching row below shows the corresponding expected average avalanche counts for each SiPM and the identified scintillator pixel (small, green rectangle). A typical pile-up event and the response of the proposed positioning algorithm are shown in the rightmost column. It can be observed that only one of the two scintillation events (the one containing the SiPM with $q_m$) is considered for position identification. Counts of identified scintillator pixels for 1147933 scintillation events and with a corrected energy between 450 keV and 600 keV are shown in fig.~\ref{subfig:xtal-counts} for the three individual scintillator pixel layers. A modulation of the counts parallel to the scanner axis (middle layer) and orthogonal to the scanner axis (uppermost layer) can be observed. The corrected energy histogram for all 1147933 classified scintillation events is shown in fig.~\ref{subfig:energy-hist-MLP}. A fit to the full absorption peak resulted in a peak position of $\mu=525.7\pm0.3$ keV, a standard deviation of $\sigma=26.4\pm0.6$ keV and an energy resolution of $\Delta E=12\%\pm2\%\, \mathrm{FWHM}$ (assuming a linear background). For comparison, the histogram for the sum of the SiPM values of the $\mathrm{3\times3}$ neighbourhood without any correction is shown in fig.~\ref{subfig:energy-hist-anger}. The energy resolution of 12\% obtained after using the scintillator pixel calibration determined by the MLP algorithm results in an energy resolution very comparable to the energy resolutions of the individual scintillator pixels shown in fig.~\ref{subfig:photopeak-eres-all-xtals}. This means that the presented algorithm and energy determination results in correct event positioning and adequate correction for varying light yields, as was the case with the iterative formulation of the MLP algorithm \parencite{Lerche2016MaximumDetectors}.

\subsection{Processing speed}

The pure MLP processing (including final scintillator pixel identification and energy calibration) 
reaches about $7.2\cdot 10^6$ singles per second using exclusively one thread (Intel Xeon Platinum 8168 CPU, 2.70GHz). The maximum singles throughput (using 60 threads), including all required processing steps (i.e. IO, unpacking, timestamp extraction and calibration, single event clustering, MLP, storing) exceeds $22.5\cdot 10^6$ singles per second. In this case, MLP only needs 
approx.~20\% of the total CPU time. I/O requires about 50\% CPU time as a major bottleneck,
and single event clustering requires 20\%, while unpacking and timestamp calculations occupy the remaining time. 

\section{Discussion}

The use of pixelated scintillator  matrices is extremely advantageous in determining the average expected values of the SiPM signals from measurements. For a nearly homogeneous gamma-photon irradiation from a 511 keV source, the pixelation automatically results in accumulation points for the specific centroids (Anger positions) associated with the corresponding scintillator pixels. This allows easy in-situ calibration of the MLP method. The calibration script for the present work took about one to two hours per scintillation detector. In contrast, monolithic scintillators require a much more complex calibration with collimated gamma-photon beams sequentially positioned at multiple defined positions to generate the required measurement data \parencite{Hunter2009CalibrationDetector,Barrett2009Maximum-likelihoodDetectors,Wenze2007MaximumExperiments}. 

 The spectra using the fit model with the corresponding start parameters are shown as grey dashed lines in Figs.~\ref{subfig:pixel-e-hists-single-peak} and \ref{subfig:pixel-e-hists-double-peak}. As stated above, energy histograms with two photopeaks (fig.~\ref{subfig:pixel-e-hists-double-peak}) can occur near the edges of the scintillation detector if the centroid for two different scintillator pixels is nearly the same due to truncation of the scintillation light distribution. In this work, the energy information was successfully used to separate the events from the two individual scintillator pixels for the calculation of the expected average signal. In Fig.~\ref{subfig:cluster-by-layer}, the cluster positions where two full absorption peaks are observed are shown with the two colours of the corresponding layers. It has been suggested as an alternative to compute the standard deviation as an estimate of the width of the signal distribution in order to distinguish between the two scintillation pixels with almost the same centroid but from different layers \parencite{Lenz2021DesignStudies}. However, this method was found to be less effective than using the total collected light as proposed in this work.
 
\begin{figure}[t!]
  \centering
 \includegraphics[interpolate=true,width=0.3\columnwidth]{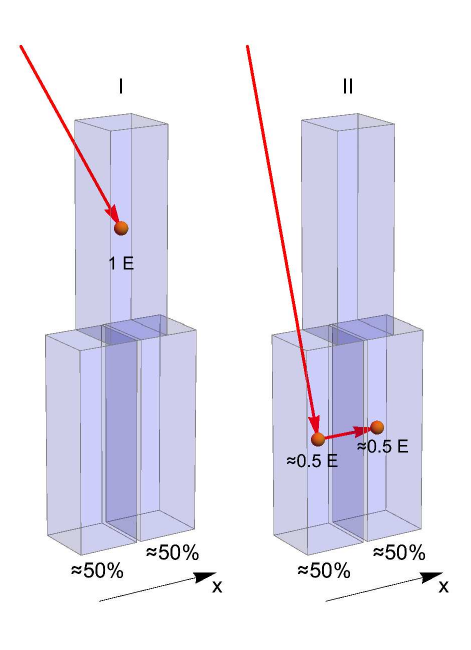}
 \caption{Two indistinguishable scintillation events. Case I: Single pixel full energy deposit in the upper pixel. Case II: ICS event with two energy deposits of about half the gamma photon energy in each of the two lower layer pixels.}
   \label{fig:ics-confusion}
\end{figure}

The small positive bias of about 3\% of the position of the full absorption peak ($\mu=525.7\pm0.3$ keV instead of $\mu=511$ keV) in the corrected energy spectrum (fig.~\ref{subfig:energy-hist-MLP}) was consistently observed for all scintillation detectors of the BrainPET 7T. Its origin is not fully understood and requires further investigation. One possible cause is contamination from pile-up events. Another possible cause is the determination of the average avalanche count distribution $\bar{q}_{i,j}$ obtained from measurements with a trigger threshold greater than zero. This leads to the censoring of avalanche count values below the threshold and consequently to an erroneous average, which could lead to a bias in the corrected energy.

The multi-threaded implementation achieves a processing speed of 22.5 million events per second, which is fully sufficient for the intended use of the BrainPET 7T insert, i.e.~multi-modal and multi-parametric imaging studies for neuroscientific questions. Typical injected activities for human studies range from about 200 MBq to 500 MBq for $\mathrm{^{18}F}$ and $\mathrm{^{15}O}$ based radiotracers, respectively. However, only about 10--20\% of the radiotracer reaches the human head. Additionally, due to the radioactive decay of the tracer during a typical scanning times (approx.~60 min), the average activity detected by the PET scanner is only about 80\% and 50\% for $\mathrm{^{18}F}$ and $\mathrm{^{15}O}$\footnote{In general, several injections are applied when using $\mathrm{^{15}O}$-based tracer.}, respectively.  
An activity of 50 MBq generates approximately 50 million detected events per second in the BrainPET 7T and approximately 180 billion detected events during one hour of scanning time. This means that position processing for the entire study can be completed within 120 minutes. 

A comparison of processing speed with most MLP methods, i.e.,  \parencite{Gray1976MaximumCameras,Fessler1991RobustCameras,Gagnon1993MaximumInteraction,Joung2000ImplementationCameras,Joung2001InvestigationCameras,Joung2001CMiCE:Scheme,Miyaoka2008DesignCapability,Wenze2007MaximumExperiments,Hesterman2010Maximum-likelihoodAlgorithm,DeWitt2010DesignPositioning,Johnson-Williams2011DesignAlgorithm,Wang2016AnDetector}, is difficult due to differences in detector design, e.g.~monolithic crystals vs.~pixelated scintillators, and due to the doubling of CPU power every 18 months \parencite{Koomey2011ImplicationsComputing}. However, the system and approach used in the study by Goldschmidt et al.~\parencite{Goldschmidt2016Software-BasedData} allow for a meaningful comparison. This study used a preclinical system with digital SiPMs, pixelated scintillation detectors and iterative MLP implementation \parencite{Lerche2016MaximumDetectors}. They reported a maximum processing speed of approximately 1.7 million single events per second with 28 MBq of activity using a multi-threaded implementation (40 threads on four $\times$ Intel Xeon X7560, $4\times 8$ cores @ 2.27 GHz). Considering that the performance of the Intel Xenon Platinum 8168 CPU is approximately 4 times higher and that 60 physical CPU cores could be used in our study compared to only 32, the processing speed with the implementation presented here is approximately two times higher, which can be attributed to the optimized implementation described in section \ref{sect:mehtods}. However, an exact comparison is not possible. While the digital SiPMs and the validation threshold are the same in both systems, their complexity is very different (3840 SiPM pixels paired with 54000 scintillator pixels in a single layer design vs.~17280 SiPM pixels paired with 196000 scintillator pixels in a triple layer design). Also, not all of the details relating to the data preprocessing are known.  
The MLP approaches presented by \cite{berker2018ScintillationEvent} and \cite{Gross-Weege2016MaximumScanners} are based on the method in \cite{Lerche2016MaximumDetectors}, but include the trigger probabilities for individual SiPM pixels \parencite{Tabacchini2014ProbabilitiesPhotomultiplier}, with a similar accuracy to the original method when comparing energy resolution and positioning performance compared to the centroid method, comparable to the original method.  

\section{Conclusions}
In this study, a fast implementation of a maximum likelihood positioning method for estimating the energy and identification of the active scintillator pixel in PET scanners with staggered layer scintillation detectors is presented. The effectiveness of the algorithm optimization was evaluated by measurements on a three-layer scintillation detector block with a total of 1634 scintillator pixels. With the iteration-free implementation and various additional optimization methods such as SIMD parallelization, multi-threading, optimized cache lines, and adapted numerical precision of the logarithm, a maximum processing speed of approximately 22.5 million single scintillation events per second was achieved using 60 threads on a platform with four Intel Xeon Platinum 8168 CPUs, where all required processing steps were included. This is approximately two times faster than the non-optimized, iterative formulation of the MLP on a comparable computing platform. The precision and accuracy of the optimized MLP is comparable to previous implementations. However, the extension of the positioning method to a three-layer scintillation detector design for depth of interaction detection revealed, that for certain scintillator pixels, single scintillation events cannot be distinguished from intercrystal scattered scintillation events. This effect is a consequence of the staggered layer design and not of the positioning algorithm.

\section{Acknowledgements}
The project was supported by the Helmholtz Validation Fund No.~0051 and the Innovation Go Funding of the Forschungszentrum Jülich GmbH for the project “BrainPET 7T”. W.~Bi received funding from Chinese Scholarship Council. Q.~Liu received financial support from the Helmholtz - OCPC Postdoctoral Fellowship Programme. E. Pfaehler is funded by the European Union, Marie-Curie Sklodowska Fellowship HORIZON-MSCA-2021-PF-01, grant 101068572. We would like to thank Claire Rick for the linguistic revision of the manuscript. 

\printbibliography

@article{Goldschmidt2016Software-BasedData,
    title = {{Software-Based Real-Time Acquisition and Processing of PET Detector Raw Data}},
    year = {2016},
    journal = {IEEE Transactions on Biomedical Engineering},
    author = {Goldschmidt, Benjamin and Schug, David and Lerche, Christoph and Salomon, Andre and Gebhardt, Pierre and Weissler, Bjoern and Wehner, Jakob and Dueppenbecker, Peter M. and Kiessling, Fabian and Schulz, Volkmar},
    number = {2},
    month = {2},
    pages = {316--327},
    volume = {63},
    url = {http://ieeexplore.ieee.org/document/7156103/},
    doi = {10.1109/TBME.2015.2456640},
    issn = {0018-9294}
}

@article{Schug2016InitialTechnology,
    title = {{Initial PET performance evaluation of a preclinical insert for PET/MRI with digital SiPM technology}},
    year = {2016},
    journal = {Physics in Medicine and Biology},
    author = {Schug, David and Lerche, Christoph and Weissler, Bjoern and Gebhardt, Pierre and Goldschmidt, Benjamin and Wehner, Jakob and Dueppenbecker, Peter Michael and Salomon, Andre and Hallen, Patrick and Kiessling, Fabian and Schulz, Volkmar},
    number = {7},
    pages = {2851--2878},
    volume = {61},
    publisher = {IOP Publishing},
    doi = {10.1088/0031-9155/61/7/2851},
    issn = {13616560},
    pmid = {26987774},
    arxivId = {1507.00536}
}

@article{Liu2013ImprovedAlgorithm,
    title = {{Improved event positioning in a gamma ray detector using an iterative position-weighted centre-of-gravity algorithm}},
    year = {2013},
    journal = {Physics in Medicine and Biology},
    author = {Liu, Chen and Goertzen, Andrew L.},
    number = {14},
    volume = {58},
    doi = {10.1088/0031-9155/58/14/N189},
    issn = {00319155},
    pmid = {23798644}
}

@article{DeWitt2010DesignPositioning,
    title = {{Design of an FPGA-based algorithm for real-time solutions of statistics-based positioning}},
    year = {2010},
    journal = {IEEE Transactions on Nuclear Science},
    author = {DeWitt, Don and Johnson-Williams, Nathan G. and Miyaoka, Robert and Li, Xiaoli and Lockhart, Cate and Lewellen, Tom K. and Hauck, Scott},
    number = {1 PART 1},
    pages = {71--77},
    volume = {57},
    doi = {10.1109/TNS.2009.2030581},
    issn = {00189499}
}

@article{Wang2016AnDetector,
    title = {{An FPGA-Based Real-Time Maximum Likelihood 3D Position Estimation for a Continuous Crystal PET Detector}},
    year = {2016},
    journal = {IEEE Transactions on Nuclear Science},
    author = {Wang, Yonggang and Xiao, Yong and Cheng, Xinyi and Li, Deng and Wang, Liwei},
    number = {1},
    pages = {37--43},
    volume = {63},
    publisher = {IEEE},
    doi = {10.1109/TNS.2015.2506739},
    issn = {00189499}
}

@article{Johnson-Williams2011DesignAlgorithm,
    title = {{Design of a real time FPGA-based three dimensional positioning algorithm}},
    year = {2011},
    journal = {IEEE Transactions on Nuclear Science},
    author = {Johnson-Williams, Nathan and Miyaoka, Robert S. and Li, Xiaoli and Lewellen, Tom K. and Hauck, Scott},
    number = {1 PART 1},
    pages = {26--33},
    volume = {58},
    publisher = {IEEE},
    doi = {10.1109/TNS.2010.2093909},
    issn = {00189499}
}

@article{Lerche2011MaximumDetectors,
    title = {{Maximum likelihood based positioning and energy correction for pixelated solid state PET detectors}},
    year = {2011},
    journal = {IEEE Nuclear Science Symposium Conference Record},
    author = {Lerche, C.W. and Solf, Torsten and Dueppenbecker, Peter and Goldschmidt, Benjamin and Marsden, Paul K. and Schulz, Volkmar},
    pages = {3610--3613},
    publisher = {IEEE},
    isbn = {9781467301183},
    doi = {10.1109/NSSMIC.2011.6153679},
    issn = {10957863}
}

@article{Lerche2009MaximumMeasurement,
    title = {{Maximum likelihood positioning for gamma-ray imaging detectors with depth of interaction measurement}},
    year = {2009},
    journal = {Nuclear Instruments and Methods in Physics Research Section A: Accelerators, Spectrometers, Detectors and Associated Equipment},
    author = {Lerche, C.W. and Ros, A. and Monz{\'{o}}, J.M. and Aliaga, R.J. and Ferrando, N. and Mart{\'{i}}nez, J.D. and Herrero, V. and Esteve, R. and Gadea, R. and Colom, R.J. and Toledo, J. and Mateo, F. and Sebasti{\'{a}}, A. and S{\'{a}}nchez, F. and Benlloch, J.M.},
    number = {1-2},
    month = {6},
    pages = {359--362},
    volume = {604},
    url = {https://linkinghub.elsevier.com/retrieve/pii/S0168900209001934},
    doi = {10.1016/j.nima.2009.01.060},
    issn = {01689002}
}

@article{Miyaoka2008DesignCapability,
    title = {{Design of a high resolution, monolithic crystal, PET/MRI detector with DOI positioning capability}},
    year = {2008},
    journal = {IEEE Nuclear Science Symposium Conference Record},
    author = {Miyaoka, Robert S. and Li, Xiaoli and Lockhart, Cate and Lewellen, Tom K.},
    pages = {4688--4692},
    isbn = {9781424427154},
    doi = {10.1109/NSSMIC.2008.4774469},
    issn = {10957863}
}

@article{Gross-Weege2016MaximumScanners,
    title = {{Maximum likelihood positioning algorithm for high-resolution PET scanners}},
    year = {2016},
    journal = {Medical Physics},
    author = {Gross-Weege, Nicolas and Schug, David and Hallen, Patrick and Schulz, Volkmar},
    number = {6},
    pages = {3049--3061},
    volume = {43},
    doi = {10.1118/1.4950719},
    issn = {24734209}
}

@article{Lerche2016MaximumDetectors,
    title = {{Maximum likelihood positioning and energy correction for scintillation detectors}},
    year = {2016},
    journal = {Physics in Medicine and Biology},
    author = {Lerche, Christoph and Salomon, André and Goldschmidt, Benjamin and Lodomez, Sarah and Weissler, Björn and Solf, Torsten},
    number = {4},
    pages = {1650--1676},
    volume = {61},
    doi = {10.1088/0031-9155/61/4/1650},
    issn = {13616560},
    pmid = {26836394}
}

@article{Ito2010ASystem,
    title = {{A four-layer DOI detector with a relative offset for use in an animal PET system}},
    year = {2010},
    journal = {IEEE Transactions on Nuclear Science},
    author = {Ito, Mikiko and Lee, Jae Sung and Kwon, Sun Il and Lee, Geon Song and Hong, Byungsik and Lee, Kyong Sei and Sim, Kwang Souk and Lee, Seok Jae and Rhee, June Tak and Hong, Seong Jong},
    number = {3 PART 1},
    pages = {976--981},
    volume = {57},
    isbn = {1424409233},
    doi = {10.1109/TNS.2010.2044892},
    issn = {00189499}
}

@article{Joung2001CMiCE:Scheme,
    title = {{cMiCE: A high resolution animal PET using continuous LSO with a statistics based positioning scheme}},
    year = {2001},
    journal = {IEEE Nuclear Science Symposium and Medical Imaging Conference},
    author = {Joung, J. and Miyaoka, R. S. and Lewellen, T. K.},
    pages = {1137--1141},
    volume = {2},
    doi = {10.1109/nssmic.2001.1009751}
}

@article{Hesterman2010Maximum-likelihoodAlgorithm,
    title = {{Maximum-likelihood estimation with a contracting-grid search algorithm}},
    year = {2010},
    journal = {IEEE Transactions on Nuclear Science},
    author = {Hesterman, Jacob and Caucci, Luca and Kupinski, Matthew A. and Barrett, Harrison H. and Furenlid, Lars R.},
    number = {3 PART 1},
    pages = {1077--1084},
    volume = {57},
    doi = {10.1109/TNS.2010.2045898},
    issn = {00189499}
}

@article{Wenze2007MaximumExperiments,
    title = {{Maximum likelihood positioning of scintillation events: Preliminary experiments}},
    year = {2007},
    journal = {IEEE Nuclear Science Symposium Conference Record},
    author = {Wenze, Xi and Seidel, Jurgen and Kakareka, John and Proffitt, James and Weisenberger, Andrew G. and Majewski, Stan and Pohida, Tom and Green, Michael V. and Choyke, Peter},
    pages = {3808--3811},
    volume = {5},
    isbn = {1424409233},
    doi = {10.1109/NSSMIC.2007.4436950},
    issn = {10957863}
}

@inproceedings{Fessler1991RobustCameras,
    title = {{Robust maximum-likelihood position estimation in scintillation cameras}},
    year = {1991},
    booktitle = {Conference Record of the 1991 IEEE Nuclear Science Symposium and Medical Imaging Conference},
    author = {Fessler, J.A. and Rogers, W.L. and Clinthorne, N.H.},
    pages = {1851--1855},
    publisher = {IEEE},
    url = {http://ieeexplore.ieee.org/document/259236/},
    isbn = {0-7803-0513-2},
    doi = {10.1109/NSSMIC.1991.259236}
}

@article{Hunter2009CalibrationDetector,
    title = {{Calibration method for ML estimation of 3D interaction position in a thick gamma-ray detector}},
    year = {2009},
    journal = {IEEE Transactions on Nuclear Science},
    author = {Hunter, William and Barrett, Harrison and Furenlid, Lars R.},
    number = {1},
    pages = {189--196},
    volume = {56},
    doi = {10.1109/TNS.2008.2010704},
    issn = {00189499}
}

@article{Joung2001InvestigationCameras,
    title = {{Investigation of bias-free positioning estimators for the scintillation cameras}},
    year = {2001},
    journal = {IEEE Transactions on Nuclear Science},
    author = {Joung, J. and Miyaoka, R. S. and Kohlmyer, S. G. and Lewellen, T. K.},
    number = {3},
    pages = {715--719},
    volume = {48},
    doi = {10.1109/23.940152},
    issn = {00189499}
}

@article{Barrett2009Maximum-likelihoodDetectors,
    title = {{Maximum-likelihood methods for processing signals from gamma-ray detectors}},
    year = {2009},
    journal = {IEEE Transactions on Nuclear Science},
    author = {Barrett, Harrison and Hunter, William and Miller, Brian William and Moore, Stephen K. and Chen, Yichun and Furenlid, Lars R.},
    number = {3},
    pages = {725--735},
    volume = {56},
    doi = {10.1109/TNS.2009.2015308},
    issn = {00189499},
    pmid = {20107527}
}

@article{Joung2000ImplementationCameras,
    title = {{Implementation of ML based positioning algorithms for scintillation cameras}},
    year = {2000},
    journal = {IEEE Transactions on Nuclear Science},
    author = {Joung, Jinhun and Miyaoka, R.S. and Kohlmyer, S. and Lewellen, T.K.},
    number = {3},
    month = {6},
    pages = {1104--1111},
    volume = {47},
    url = {https://ieeexplore.ieee.org/document/856555/},
    doi = {10.1109/23.856555},
    issn = {0018-9499}
}

@article{Gray1976MaximumCameras,
    title = {{Maximum a Posteriori Estimation of Position in Scintillation Cameras}},
    year = {1976},
    journal = {IEEE Transactions on Nuclear Science},
    author = {Gray, Robert and Macovski, Albert},
    number = {1},
    pages = {849--852},
    volume = {23},
    url = {http://ieeexplore.ieee.org/document/4328354/},
    doi = {10.1109/TNS.1976.4328354},
    issn = {0018-9499}
}

@article{Gagnon1993MaximumInteraction,
    title = {{Maximum likelihood positioning in the scintillation camera using depth of interaction}},
    year = {1993},
    journal = {IEEE Transactions on Medical Imaging},
    author = {Gagnon, D. and Pouliot, N. and Laperriere, L. and Therrien, M. and Olivier, P.},
    number = {1},
    month = {3},
    pages = {101--107},
    volume = {12},
    url = {http://ieeexplore.ieee.org/document/222673/},
    doi = {10.1109/42.222673},
    issn = {02780062}
}

@inproceedings{Lerche2023FirstNeuroscience,
    title = {{First Performance Results of a UHF-MRI Compatible BrainPET Insert for Neuroscience}},
    year = {2023},
    booktitle = {2023 IEEE Nuclear Science Symposium, Medical Imaging Conference and International Symposium on Room-Temperature Semiconductor Detectors (NSS MIC RTSD)},
    author = {Lerche, C.W. and Niek{\"{a}}mper, D. and Scheins, J. J. and Tellmann, L. and Ridder, D. and Kitten, Y. and Skudarnov, P. and Choi, C.-H. and Felder, J. and Arutinov, D. and Heil, R. and Silex, W. and Scherer, B. and van Waasen, S. and Grunwald, D. and Lennartz, M. and Natour, G. and Weissler, B. and M{\"{u}}ller, F. and Schug, D. and Gegenmantel, E. and Radermacher, H. and Gebhardt, P. and Lefaucheur, J. L. and Chen, Z. and Egan, G. and Schulz, V. and Shah, N. J.},
    month = {11},
    pages = {1--2},
    publisher = {IEEE},
    url = {https://ieeexplore.ieee.org/document/10338309/},
    isbn = {979-8-3503-3866-9},
    doi = {10.1109/NSSMICRTSD49126.2023.10338309}
}

@inproceedings{Lerche2024MRApplications,
    title = {{MR Compatibility Evaluation of a BrainPET Insert for UHF-MR Systems and Neuroscientific Applications}},
    year = {2024},
    booktitle = {2024 IEEE Nuclear Science Symposium (NSS), Medical Imaging Conference (MIC) and Room Temperature Semiconductor Detector Conference (RTSD)},
    author = {Lerche, C.W. and Niek{\"{a}}mper, D. and Scheins, J. J. and Tellmann, L. and Choi, C.-H. and Felder, J. and Arutinov, D. and Heil, R. and Silex, W. and Van Waasen, S. and Grunwald, D. and Lennartz, M. and Meurere, T. and Natour, G. and Wei{\ss}ler, B. and M{\"{u}}ller, F. and Schug, D. and Gegenmantel, E. and Radermacher, H. and Gebhardt, P. and Lefaucheur, J. L. and Chen, Z. and Egan, G. and Schulz, V. and Shah, J.},
    month = {10},
    pages = {1--1},
    publisher = {IEEE},
    url = {https://ieeexplore.ieee.org/document/10656041/},
    isbn = {979-8-3503-8815-2},
    doi = {10.1109/NSS/MIC/RTSD57108.2024.10656041}
}

@article{Du1999CentroidalAlgorithms,
    title = {{Centroidal Voronoi tessellations: Applications and algorithms}},
    year = {1999},
    journal = {SIAM Review},
    author = {Du, Qiang and Faber, Vance and Gunzburger, Max},
    number = {4},
    pages = {637--676},
    volume = {41},
    doi = {10.1137/S0036144599352836},
    issn = {00361445}
}

@article{Tabacchini2014ProbabilitiesPhotomultiplier,
    title = {{Probabilities of triggering and validation in a digital silicon photomultiplier}},
    year = {2014},
    journal = {Journal of Instrumentation},
    author = {Tabacchini, V. and Westerwoudt, V. and Borghi, G. and Seifert, S. and Schaart, D. R.},
    number = {6},
    volume = {9},
    doi = {10.1088/1748-0221/9/06/P06016},
    issn = {17480221}
}

@phdthesis{Lenz2021DesignStudies,
    title = {{Design and characterisation of an MRI compatible human brain PET insert by means of simulation and experimental studies}},
    year = {2021},
    author = {Lenz, Mirjam},
    school = {Bergische Universit{\"{a}}t Wuppertal},
    doi = {10.25926/d4s5-bc37}
}

@article{Frach2009ThePerformance,
    title = {{The digital silicon photomultiplier - Principle of operation and intrinsic detector performance}},
    year = {2009},
    journal = {IEEE Nuclear Science Symposium Conference Record},
    author = {Frach, Thomas and Prescher, Gordian and Degenhardt, Carsten and De Gruyter, Rik and Schmitz, Anja and Ballizany, Rob},
    pages = {1959--1965},
    publisher = {IEEE},
    isbn = {9781424439621},
    doi = {10.1109/NSSMIC.2009.5402143},
    issn = {10957863}
}

@inproceedings{Weissler2020HyperionPET/MRI,
    title = {{Hyperion III – A flexible PET detector platform for simultaneous PET/MRI}},
    year = {2020},
    author = {Weissler, B and Schug, D and Dey, T and Gebhardt, P and Krueger, K and Mueller, F and Radermacher, H and Gross-Weege, N and Yin, L and Nadig, V and Schulz, V},
    month = {4},
    url = {http://www.thieme-connect.de/DOI/DOI?10.1055/s-0040-1708245},
    doi = {10.1055/s-0040-1708245}
}

@article{VanDam2011ImprovedDetectors,
    title = {{Improved nearest neighbor methods for gamma photon interaction position determination in monolithic scintillator PET detectors}},
    year = {2011},
    journal = {IEEE Transactions on Nuclear Science},
    author = {Van Dam, Herman T. and Seifert, Stefan and Vinke, Ruud and Dendooven, Peter and L{\"{o}}hner, Herbert and Beekman, Freek J. and Schaart, Dennis R.},
    number = {5 PART 1},
    pages = {2139--2147},
    volume = {58},
    doi = {10.1109/TNS.2011.2150762},
    issn = {00189499}
}

@phdthesis{Lodomez2012DevelopmentPET-detectors,
    title = {{Development and evaluation of crystal identification algorithms for high resolution PET-detectors}},
    year = {2012},
    author = {Lodomez, Sarah},
    school = {Rheinisch-Westf{\"{a}}lische Technische Hochschule Aachen}
}

@article{Koomey2011ImplicationsComputing,
    title = {{Implications of historical trends in the electrical efficiency of computing}},
    year = {2011},
    journal = {IEEE Annals of the History of Computing},
    author = {Koomey, Jonathan and Berard, Stephen and Sanchez, Marla and Wong, Henry},
    number = {3},
    pages = {46--54},
    volume = {33},
    publisher = {IEEE},
    doi = {10.1109/MAHC.2010.28},
    issn = {10586180}
}

@misc{Lerche2014PixelArrays,
    title = {{Pixel identification for small pitch scintillation crystal arrays}},
    year = {2014},
    author = {Lerche, Christoph and Sampath, Poornima and Solf, Torsten},
    url = {https://worldwide.espacenet.com/patent/search/family/049000310/publication/US9753146B2?q=pn%3DUS9753146B2}
}

@phdthesis{Sampath2013CharacterizationApplications,
    title = {{Characterization of High Performance Gamma Ray Imaging Detectors for Digital PET in Preclinical and Clinical Applications}},
    year = {2013},
    author = {Sampath, Poornima},
    pages = {1--99},
    school = {Aachen University of Applied Sciences}
}

@misc{berker2018ScintillationEvent,
  title={{Scintillation event position determination in a radiation particle detector}},
  author={Berker, Yannick and  Schulz, Volkmar},
  year={2018},
  note={US9903960B1},
  month={02},
  day={27},
  howpublished={Espacenet},
  url ={https://worldwide.espacenet.com/patent/search/family/052692491/publication/WO2016146391A1?q=pn%3DWO2016146391A1}
}

@misc{mineiro2011FasApproximate,
  title={{Fast Approximate Logarithm, Exponential, Power, and Inverse Root}},
  author={Mineiro, Paul},
  year={2011},
  month={06},
  day={20},
  howpublished={Machined Learnings},
  url ={http://www.machinedlearnings.com/2011/06/fast-approximate-logarithm-exponential.html}
}

@misc{Lerche2019PatentMethodDetectors,
    title = {{Method for the position and energy determination in scintillation detectors}},
    year = {2019},
    author = {Lerche, C. and Bi, W. and Shah, N.J.},
    institution = {Forschungszentrum Juelich GmbH},
    howpublished={Espacenet},
    url ={https://worldwide.espacenet.com/patent/search/family/072518272/publication/DE102019007136B3?q=pn%3DDE102019007136B3}
}

@misc{lodomez2014AparatusEvaluation,
  title={{Apparatus and method for the evaluation of gamma radiation events}},
  author={Lerche, C.W. and Lodomez, S. and Schulz, V. and Weissler, B.},
  year={2014},
  note={EP2994776B1},
  month={04},
  day={30},
  howpublished={Espacenet},
  url ={https://worldwide.espacenet.com/patent/search/family/048444081/publication/EP2994776B1?q=pn%3DEP2994776B1}
}

@misc{intel2021, 
    title = {{Intel C++ Compiler Classic Developer Guide and Reference}},
    author = {Intel},
    year = {2021},
    url ={https://www.intel.com/content/www/us/en/docs/cpp-compiler/developer-guide-reference/2021-9/intrinsics-for-avx-512-instructions.html}
}

@book{Kowarschik2003,
    author = {Kowarschik, M. and Weiss, C.},
    title = {{An overview of cache optimization techniques and cache-aware numerical algorithms Algorithms for Memory Hierarchies—Advanced Lectures}},
    publisher ={Springer, Berlin} ,
    year = {2003} 
}
\end{document}